\documentclass[dvips]{article}

\usepackage{icrc2011}

\title{Testing the emission models of blazar jets with the MAGIC telescopes}

\newcommand{\etal}{\MakeLowercase{\textit{et al. }}} 
\shorttitle{J.~Becerra-Gonz\'alez \etal Testing the emission models of blazar jets}

\authors{J.~Becerra-Gonz\'alez$^{1,2}$, A.~Stamerra$^{3}$, K.~Saito$^{4}$, D.~Mazin$^{5}$, F.~Tavecchio$^{6}$, L.~Maraschi$^{6}$, E.~Prandini$^{7}$, J.~Sitarek$^{8}$ and K.Berger$^{1}$ on behalf of the MAGIC collaboration.}
\afiliations{$^1$ Instituto de Astrof\'isica de Canarias\\ $^2$ Universidad de La Laguna\\$^3$ Universita di Siena and INFN Pisa\\$^4$ Max-Planck-Institut fur Physik\\$^5$ Institut de Fisica d'Altes Energies, Barcelona\\$^6$ INAF National Institute for Astrophysics \\$^7$ Universita di Padova and INFN\\$^8$ University of Lodz  }
\email{jbecerra@iac.es}

\abstract{The MAGIC telescopes discovered very high energy (VHE, E$>$100~GeV) gamma-ray emission
coming from the distant Flat Spectrum Radio Quasar (FSRQ) PKS 1222+21 (4C +21.35,
z=0.432). It is the second most distant VHE gamma-ray source, with well measured
redshift, detected until now. The observation was performed on 2010 June 17 (MJD~55364.9)
using the two 17 m diameter imaging Cherenkov telescopes on La Palma (Canary Islands,
Spain). The MAGIC detection coincides with high energy MeV/GeV gamma-ray activity
measured by the Large Area Telescope (LAT) on board the Fermi satellite. The averaged
integral flux above 100 GeV is equivalent to 1 Crab Nebula flux. The VHE flux
measured by MAGIC varies significantly within the 30~minutes of exposure implying a flux doubling time of about 10~minutes. The
VHE and MeV/GeV spectra, corrected for the absorption by the extragalactic background
light, can be described by a single power law with photon index $2.72\pm0.34$ between 3~GeV
and 400~GeV, consistent with gamma-ray emission belonging to a single component in the
jet. The absence of a spectral cutoff at 30-60~GeV (indeed, one finds a strict lower
limit $E_c>$~130~GeV) constrains the gamma-ray emission region to lie outside the
broad line region, which would otherwise absorb the VHE gamma-rays. Together with the
detected fast variability, this challenges present emission models from jets in FSRQs.}

\begin{document}
\maketitle

\section{Introduction}

Blazars are active galactic nuclei hosting  powerful relativistic jets pointing toward the observer. They are characterized by strong non-thermal emission extending across the entire electromagnetic spectrum, from radio up to $\gamma$-rays. The blazar PKS~1222+21 (also knows as 4C~21.35) belongs the class of Flat Spectrum Radio Quasars (FSRQs) and is located at a redshift of $z=0.432$~\cite{Osterbrock1987}. FSRQs display luminous broad emission lines, often accompanied by a ``blue bump" in the optical-UV region which is associated with the direct emission from the accretion disk. This optical-UV emission is believed to be re-processed by the Broad Line Region (BLR) clouds, filling the BLR with a dense optical-UV radiation field which can interact with VHE $\gamma$-rays causing internal absorption in the VHE domain~\cite{2003APh....18..377D}.
PKS~1222+21 is the third FSRQ detected in VHE $\gamma$-rays after PKS~1510-08~\cite{Wagner2010} and 3C~279~\cite{MAGIC3c279, 3C279magic2011} at the redshifts of $z=0.36$ and $z=0.536$, respectively. It is thus the second most distant VHE source after 3C~279 with well measured redshift (the BL~Lac 3C~66A with an estimated redshift $z=0.444$~\cite{2005ApJ...629..108B} would occupy this position of the ranking, but the measurement has a large uncertainty).

In this proceeding, we will present the MAGIC discovery of this source which was detected during a $\gamma$-rays flare announced by the \textit{Fermi}/LAT collaboration and its physics implications.

\section{MAGIC discovery}
MAGIC consists of two 17\,m diameter Imaging Atmospheric Cherenkov Telescopes (IACTs). The data were taken at zenith angles between $26^\circ$ and $35^\circ$. Stereoscopic events, triggered by both telescopes, were analyzed in the MARS analysis framework~\cite{MoralejoLodz}. Details on the analysis can be found in \cite{3C66Amagic2010} whereas the performance of the MAGIC stereo system will be discussed in detail in a forthcoming paper \cite{Stereoinprep}. 

\begin{figure}[!t]
  \vspace{5mm}
  \centering
  \includegraphics[width=3.in]{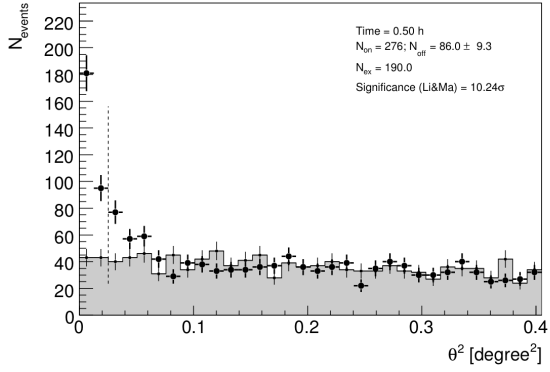}
\caption{\centering{\small{Distribution of the squared angular distance ($\theta^2$)  for events in the direction of 4C +21.35 (black points) and normalized off-source events (grey histogram). The signal is extracted in the $\theta^2$-region denoted by the vertical dashed line.}}}
\label{fig:theta2}
 \end{figure}

PKS~1222+21 was observed by MAGIC on June 17 (MJD 55364) for 30~minutes as  part of a Target of Opportunity program triggered by an increased flux in the \textit{Fermi} energy band~\cite{FermiATel}. During this detection by MAGIC the source was close to the brightest flare ever observed by the \textit{Fermi} Large Area Telescope (LAT)~\cite{Fermi draft}. The signal evaluation was performed using the $\theta^2$ distribution (squared angular distance between the true and reconstructed source position), see Fig.~\ref{fig:theta2}. We got an excess of 190 $\gamma$-like events (6\,$\gamma$/min.) above a background of 86 events, which corresponds to a statistical significance of 10.2\,$\sigma$ using eq.\ 17 in \cite{LiMa}. The energy threshold of this analysis is $\approx 70\,$GeV.

\subsection{Very High Energy Spectrum}

\begin{figure}[!t]
  \vspace{5mm}
  \centering
  \includegraphics[width=3.in]{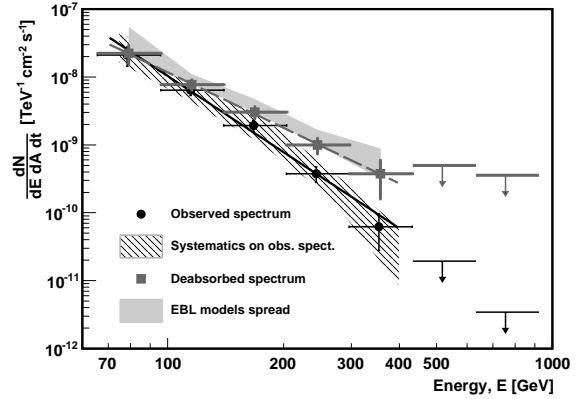}
\caption{\small{Differential energy spectrum of PKS~1222+21 as 
measured by MAGIC on 2010 June 17. Differential fluxes are shown as black points, 
upper limits (95\% C.L.)  as  black arrows. The black line is the best fit to a power law.
The black-striped area represents the systematic uncertainties of the analysis. 
The absorption corrected spectrum and upper limits using the EBL model of
\cite{Dominguez2010} are shown by the grey squares and arrows; the dashed grey line is the best fit power law.
The grey shaded area illustrates the uncertainties  due to differences in the EBL models of \cite{kneiskedole2010,gilmore:2009a,franceschini} and \cite{MAGIC3c279}.}}
\label{fig:spectrum}
 \end{figure}

The differential energy spectrum measured by MAGIC extends from 70\,GeV to 400\,GeV (Figure~\ref{fig:spectrum}) and it is well-described by a simple power law of the form:
\begin{equation}
\frac{dN}{dE}=N_{200}\left(\frac{E}{200\,\mbox{GeV}}\right)^{-\Gamma}
\end{equation}
with a photon index $\Gamma~=~3.75~\pm0.27_{\tiny\mbox{stat}}\pm0.2_{\tiny\mbox{syst}}$ and a normalization constant at 200\,GeV of 
$N_{\tiny 200}~=~(7.8~\pm~1.2_{\tiny\mbox{stat}}\pm3.5_{\tiny\mbox{syst}})\times10^{-10}\mbox{cm}^{-2}\mbox{s}^{-1}  \mbox{TeV}^{-1}$, yielding an integral flux $(4.6\pm0.5)\times 10^{-10}\,\mbox{cm}^{-2}\mbox{s}^{-1}$ ($\approx 1$ Crab Nebula flux) at E$>100\,$GeV. For energies higher than 400\,GeV no significant excess was found but two upper limits corresponding to 95\% confidence level (C.L.) have been derived. The method used for the spectral reconstruction is the so-called  ``Tikhonov" unfolding algorithm \cite{unfolding}, which takes into account the finite energy resolution of the instrument and the biases in the energy reconstruction.
The systematic uncertainty of the analysis (studied by using different cuts and different unfolding algorithms) is shown by the striped area.

The deabsorbed spectrum is shown by the grey squares in Fig.~\ref{fig:spectrum}, where the EBL model of  \cite{Dominguez2010} has been used. It is well fitted  by a power law with an intrinsic photon index of $\Gamma_{\tiny\mbox{intr}}~=~2.72~\pm0.34$ between 70\,GeV and 400\,GeV. Uncertainties caused by the differences between the EBL models are represented in Fig.~\ref{fig:spectrum} by the grey shaded area. The corresponding spread is smaller than the systematic uncertainties of the MAGIC data analysis.

Using the $\chi^2$ difference method we studied the possibility of a cut-off in the energy spectrum. We tried to fit the spectrum by a broken power law with different photon indexes and values for the cut-off. We concluded from this study that with the available statistics, at the 95\% C.L. we cannot exclude the presence of a cut-off above 130 GeV for a photon index 2.4 (the lowest possible value compatible with fit uncertainties and with the \textit{Fermi}/LAT data, see Fig.~\ref{fig:SED}) or above 180 GeV for a photon index 2.7.

\subsection{Variability in Very High Energy $\gamma$-rays}

\begin{figure}[!t]
  \vspace{5mm}
  \centering
  \includegraphics[width=3.in]{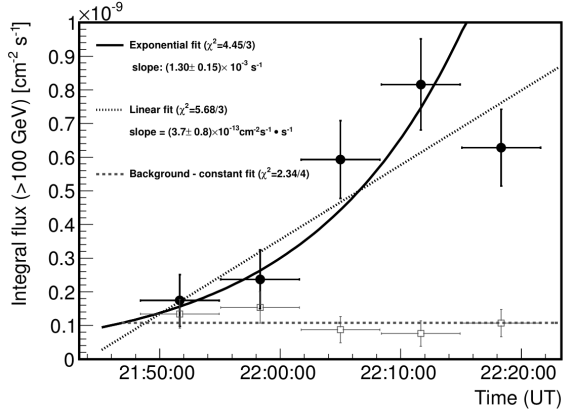}
\caption{\small{PKS~1222+21 light curve above 100\,GeV, in 6 minute bins (black filled circles). The observation was carried out on MJD 55364.  The black solid line is a fit with an exponential function and the black dotted line a fit with a linear function. The grey open squares denote the fluxes from the background events and the grey 	dashed line is a fit with a constant function to these points.}}
\label{fig:LC}
 \end{figure}

Thanks to the strength of the signal even if the observation time is as short as 30 minutes, a variability study is possible. In Fig.~\ref{fig:LC} the light curve binned in 6 minute long intervals is shown. It reveals a flux variation within the 30 minutes of observation time. The hypothesis of a constant flux is rejected ($\chi ^2/{NDF} = 28.3/4$) with high confidence (probability $<1.1\times 10^{-5}$). The flux of the background events surviving the $\gamma$/hadron selection cuts is compatible with being constant and hence we can exclude a variation of the instrument performance during the observation.

To quantify the variability time scale we performed an exponential fit (solid black line  in Fig.~\ref{fig:LC}). A linear fit is also acceptable but does not allow us to define a time scale unambiguously. For the exponential fit  the doubling time of the flare is estimated  as $8.6^{+1.1}_{-0.9}$\,minutes. The derived timescale corresponds to the fastest time variation ever observed in an FSRQ in the VHE range and in any other energy range~\cite{Foschini2011}, and is amongst the shortest measured from any TeV emitting source~\cite{HESS2155(2010),Mkr501}.

\section{Spectral Energy Distribution}

\begin{figure}[!t]
  \vspace{5mm}
  \centering
  \includegraphics[width=3.in]{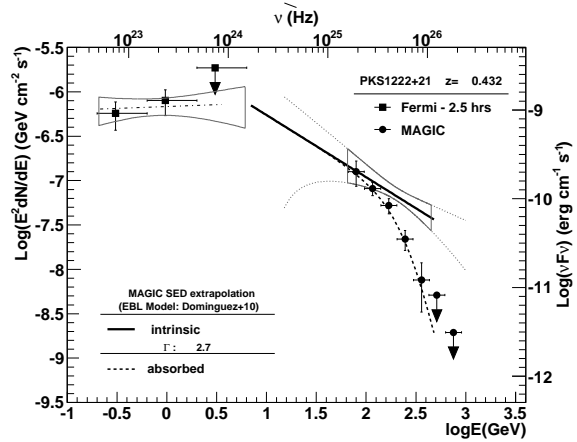}
\caption{\small{High energy SED of PKS~1222+21 during the flare of 2010 June 17 (MJD 55364.9), showing  \textit{Fermi}/LAT (squares) and MAGIC (circles) differential fluxes. 
A red bow tie in the MeV/GeV range represents the uncertainty of the likelihood fit to the \textit{Fermi}/LAT data.
The unfolded and deabsorbed spectral fit of the MAGIC data is also shown as a grey bow tie, extrapolated to lower and higher energies (dotted lines). A thick solid line (photon index $\Gamma=2.7$) 
indicates a possible extrapolation of the MAGIC deabsorbed data to lower energies. 
The thick dashed line represents the EBL absorbed spectrum obtained from the extrapolated intrinsic spectrum using  the model of \cite{Dominguez2010}. 
}}
\label{fig:SED}
 \end{figure}

The high-energy Spectral Energy Distribution (SED) is shown in Figure \ref{fig:SED} where the MeV/GeV energy range spectrum from \textit{Fermi}/LAT and the GeV/TeV spectrum measured by MAGIC are combined. The source showed a significant flare in the \emph{Fermi} band lasting $\sim$3 days, with a peak flux on 2010 June 18 (MJD 55365)~\cite{Fermi draft}.  During the 30~minutes MAGIC observation there is a gap in the LAT exposure, so we analyzed the closest data available, a period of  2.5\,hr (MJD 55364.867 to 55364.973) before and after the MAGIC observation. Given the short observation time (chosen in order to be as much contemporaneous as possible with the MAGIC data) there is no detection above 2\,GeV in the  \textit{Fermi}/LAT data, but an upper limit at the 95\% C.L. in the energy range $2-6.3$\,GeV has been calculated together with the spectral points up to 2\ GeV and combined with the MAGIC data in the SED shown in  Fig.~\ref{fig:SED}.

If we extrapolate the intrinsic MAGIC spectrum to lower energies we can see that there is a potentially smooth connection between the  \textit{Fermi}/LAT and MAGIC  extrapolated data and  the photon index steepens from 1.9 in the  \textit{Fermi}/LAT range to 2.7 in the MAGIC range. These results agree with the analysis of larger temporal intervals during this flare and during the whole active period, in which the source spectrum is well described by a broken power law with an energy break between 1 and 3 GeV~\cite{Fermi draft}. Furthermore it is found that the high energy tail  (E\,$>2\,$GeV) of the \textit{Fermi}/LAT spectrum of 4C +21.35 extends up to 50 GeV, with a photon index in the range 2.4-2.8. 

\section{Physical implications on the jet emission models}

As we have discussed, the almost simultaneous VHE and GeV spectra are consistent with a single power law with index $\sim 2.7\pm0.3$ between 3~GeV and 400~GeV, without a strong intrinsic cutoff and with a smooth connection to the lower energy spectrum. This evidence suggests that the 100~MeV - 400~GeV emission belongs to a unique component, peaking at  $\approx 2-3\,$GeV, produced in a single region of the jet. 
Considering the inverse Compton scattering on external photons and relativistic electrons in the jet as the emission process, as it is usually assumed, we have two possible scenarios. The emitting jet region could be inside of the Broad Line Region (BLR), where the external photons field would be the UV photons from the BLR or we can  assume this emitting region to be outside of the BLR where the photon field would be composed of the IR photons coming from the torus. There are two important effects that need to be taken into account:  the decreased  efficiency of the IC scattering occuring in  the Klein-Nishina (KN) regime and the absorption of $\gamma$-rays through pair production. 

The energy above which the KN effects and $\gamma$-ray absorption become important in the case of UV external photons from the BLR is $\sim$~tens of GeV, and thus we would expect a cut-off at these energies if the emitting jet region is inside the BLR. But if we consider as target photons the ones coming from the IR torus this effect would start to be appreciated only at much higher energies above $\approx$~1\,TeV. Since there is no evidences of a cut-off at low energies we can conclude that the emission should come from outside of the BLR, as has been proposed by the ``far dissipation" scenarios [e.g. \cite{Sikora2008}].


Besides, the other important result of the MAGIC observation is the fast variability, $t_{\rm var}\sim10$~minutes which indicates an extremely compact emission region. This is difficult to reconcile with the ``far dissipation" scenarios if the emission takes place in the entire cross section of the jet since in this case the emitting region should be close to the central black hole and thus inside of the BLR.  

Some explanations have been already proposed in order to solve this kind of incongruities, invoking the presence of very compact emission regions embedded within the large scale jet \cite{GhiselliniTavecchio2008,blob} or the possibility of a very strong jet recollimation [e.g. \cite{NalewajkoSikora2009}].

In conclusion the MAGIC observation of VHE emission from the FSRQ PKS~1222+21 sets severe constraints on the emission models of blazar jets. 

\section{Acknowledgments}

We would like to thank the Instituto de Astrof\'{\i}sica de
Canarias for the excellent working conditions at the
Observatorio del Roque de los Muchachos in La Palma.
The support of the German BMBF and MPG, the Italian INFN, 
the Swiss National Fund SNF, and the Spanish MICINN is 
gratefully acknowledged. This work was also supported by 
the Marie Curie program, by the CPAN CSD2007-00042 and MultiDark
CSD2009-00064 projects of the Spanish Consolider-Ingenio 2010
programme, by grant DO02-353 of the Bulgarian NSF, by grant 127740 of 
the Academy of Finland, by the YIP of the Helmholtz Gemeinschaft, 
by the DFG Cluster of Excellence ``Origin and Structure of the 
Universe'', and by the Polish MNiSzW grant 745/N-HESS-MAGIC/2010/0.

\clearpage


\begin{thebibliography}{}
\bibitem{Osterbrock1987}{D.~E.~Osterbrock \& R.~W.~Pogge, ApJ,1987, {\bf 323}, 108}
\bibitem{Wagner2010}{Wagner, S. \& Behera, B. ,10th HEAD Meeting, Hawaii, 2010, BAAS 42, {\bf 2}, 07.05}
\bibitem{MAGIC3c279}{J.~Albert et al., Science,2008, {\bf 320}, 1752}
\bibitem{3C279magic2011}{J.~Aleksic et al., A\&A, 2011, {\bf 530}, A4}
\bibitem{2005ApJ...629..108B}{D.A. Bramel et al., ApJ, 2005, {\bf 629}, 108-114}
\bibitem{2003APh....18..377D}{A.-C. Donea \& R. J. Protheroe, Astroparticle Physics, 2003, {\bf 18}, 377-393}
\bibitem{LiMa}{T.-P.~Li \& Y.-Q.~Ma, ApJ, 1983, {\bf 272}, 317}
\bibitem{MoralejoLodz}{A.~Moralejo et al., Proc. of 31st ICRC, 2009, {arXiv:0907.0943}}
\bibitem{FermiATel}{D.~Donato, The Astronomer's Telegram, \#2584}
\bibitem{Stereoinprep}{J.~Aleksic et al., \emph{in prep.}}
\bibitem{Dominguez2010}{A.~Dominguez et al., MNRAS, 2011, {\bf 410}, 2556}
\bibitem{kneiskedole2010}{T.~M.~Kneiske \& H.~Dole, A\&A, 2010, {\bf 515}, 19}
\bibitem{gilmore:2009a}{R.~Gilmore et al., MNRAS, 2009, {\bf 399}, 1694}
\bibitem{franceschini}{A.~Franceschini et al., A\&A, 2008, {\bf 487}, 837}
\bibitem{unfolding}{J.~Albert et al., Nucl. Instr. Meth., 2007, {\bf A 583}, 494}
\bibitem{NalewajkoSikora2009}{K.~Nalewajko \& M.~Sikora, MNRAS, 2009, {\bf 392}, 1205}
\bibitem{Mkr501}{J.~Albert et al., Physics Letters B, 2008, {\bf 668}, 253 [{arXiv:0708.2889v3}]}
\bibitem{GhiselliniTavecchio2008}{G.~Ghisellini \& F.~Tavecchio, MNRAS, 2008, {\bf  386}, 28}
\bibitem{Sikora2008}{M.~Sikora et a., ApJ, 2008, {\bf 675}, 71}
\bibitem{Foschini2011}{L.~Foschini et al., A\&A, 2011, {bf 530}, A77}
\bibitem{HESS2155(2010)}{A.~Abramowski et al., A\&A, 2010, {\bf 520}, 83}
\bibitem{Fermi draft}{Y.T.~Tanaka, ApJ, 2011, {\bf 733} (1), 19}
\bibitem{3C66Amagic2010}{J.~Aleksic et al., ApJL, 2010, {\bf 726}, 58 [{arxiv:1010.0550}]}
\bibitem{blob}{F.~Tavecchio et al., Submitted to A\&A, 2011,arXiv:1104.0048v1}
\end{thebibliography}
\end{document}